\documentclass[runningheads]{llncs}

\usepackage{eccv}

\usepackage{graphicx}
\usepackage{multirow}
\usepackage{tikz}
\usepackage{comment}
\usepackage{amsmath,amssymb} 
\usepackage{color,xcolor,colortbl}

\definecolor{cvprblue}{rgb}{0.21,0.49,0.74}
\usepackage[pagebackref,breaklinks=true,colorlinks,bookmarks=false,citecolor=cvprblue,linkcolor=cvprblue]{hyperref}

\usepackage{caption}
\usepackage{wrapfig}
\usepackage{placeins}
\usepackage{booktabs}

\usepackage{eccvabbrv}

\usepackage{graphicx}
\usepackage{booktabs}

\usepackage[accsupp]{axessibility}  

\usepackage{hyperref}

\usepackage{bm}
\newcommand{\ul}[1]{\underline{\smash{#1}}}
\definecolor{lightgray}{gray}{0.95}
\definecolor{color3}{gray}{0.95}
\definecolor{rouse}{rgb}{0.981,0.961,0.941}

\begin{document}

\title{Radiative Gaussian Splatting \\ for Efficient X-ray Novel View Synthesis} 

\author{Yuanhao Cai $^{1}$, Yixun Liang $^{2}$, Jiahao Wang $^{1}$, Angtian Wang $^{1}$, \\  Yulun Zhang $^{3,\dagger}$, Xiaokang Yang $^3$, Zongwei Zhou $^{1,\dagger}$, and Alan Yuille $^{1}$ \\
}
\authorrunning{Yuanhao Cai \emph{et al.}}

\institute{$^{1}$ Johns Hopkins University, $^2$ HKUST(GZ), $^3$ Shanghai Jiao Tong University
}

\maketitle

\vspace{-4.5mm}
\begin{abstract}
X-ray is widely applied for transmission imaging due to its stronger penetration than natural light. When rendering novel view X-ray projections, existing methods mainly based on NeRF suffer from long training time and slow inference speed. In this paper, we propose a 3D Gaussian splatting-based method, namely X-Gaussian, for X-ray novel view synthesis and direct CT reconstruction. Firstly, we redesign a radiative Gaussian point cloud model inspired by the isotropic nature of X-ray imaging. Our model excludes the influence of view direction when learning to predict the radiation intensity of 3D points. Based on this model, we develop a Differentiable Radiative Rasterization (DRR) with CUDA implementation. Secondly, we formulate a principled Radiative Gaussian Volume Querying (RGVQ) method that defines a continuous attenuation field from the original radiative point attributes and enables direct CT volume reconstruction through CUDA-based parallel querying, extending conventional 3DGS beyond image rendering. Thirdly, we customize an Angle-pose Cuboid Uniform Initialization (ACUI) strategy that directly uses the parameters of the X-ray scanner to compute the camera information and then uniformly samples point positions within a cuboid enclosing the scanned object. On X-ray novel view synthesis, experiments show that our X-Gaussian outperforms state-of-the-art (SOTA) methods by \textbf{6.5 dB} while enjoying less than \textbf{15\%} training time and over \textbf{73$\bf \times$} inference speed. For direct CT reconstruction, X-Gaussian outperforms SOTA methods by \textbf{0.40 dB} and \textbf{4.34 dB} in average 3D PSNR on synthetic and real scenes, respectively, while achieving \textbf{4.30$\times$} faster inference of the learned CT volume on average. Code, models, and datasets are at \url{https://github.com/caiyuanhao1998/X-Gaussian}
\end{abstract}

\let\thefootnote\relax\footnotetext{$\dagger =$ corresponding authors.}

\vspace{-6mm}
\section{Introduction}
\vspace{-1.5mm}
X-ray novel view synthesis (NVS) aims to create X-ray projections of an object from new viewpoints that are not originally captured, using only existing projections scanned from different view directions. As we know, X-ray has stronger penetrating power to capture internal structures of imaged objects and is therefore widely applied in medical imaging~\cite{x_ray_1,x_ray_2,x_ray_3,x_ray_4,cbct_3,iterative_1}. Nonetheless, X-ray is harmful to human body due to its powerful ionizing radiation, especially when the dose of X-ray increases. Improving NVS techniques can help reduce the exposure to X-rays and provide comprehensive viewpoints of imaged parts for doctors and downstream tasks such as sparse-view CT reconstruction. Hence, X-ray NVS is very important and valuable. We study this task in the circular cone beam X-ray scanning scenario~\cite{cbct,mv_x_2,mv_x_1,mv_x_3,analytical_2,vw_art,iterative_2,iterative_3}. 

\begin{figure*}[t]
	\begin{center}
		\begin{tabular}[t]{c}  \hspace{-3.8mm}
			\includegraphics[width=1.00\textwidth]{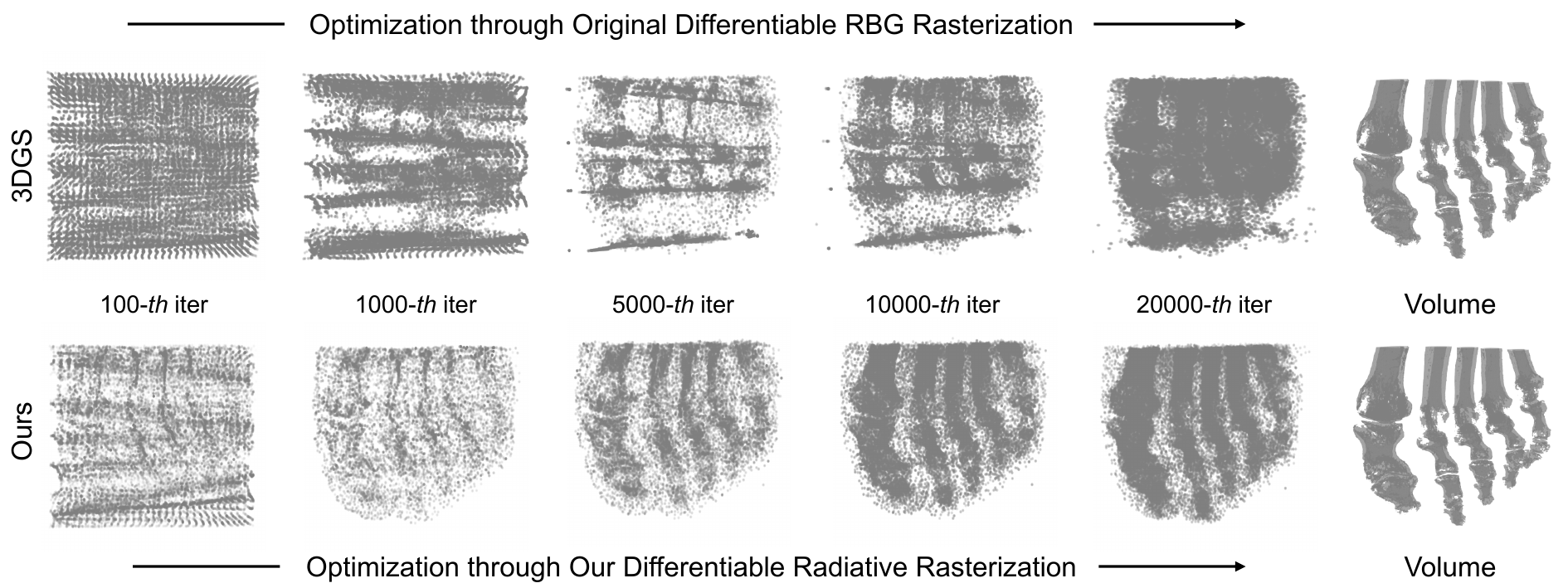}
		\end{tabular}
	\end{center}
    \vspace{-3mm}
	\caption{Point cloud visualization of the original 3DGS~\cite{3dgs} (top) and our X-Gaussian (bottom). We visualize the positions and opacities of the Gaussian point clouds at different training iterations. We also visualize the volume of foot as a reference. Note that the volume is not the ground truth of point clouds. Our X-Gaussian can better represent the detailed structures than 3DGS, showing faster and better convergence. }
    \vspace{-3mm}
	\label{fig:analysis_3dgs}
\end{figure*}

Existing methods are mainly based on neural radiance fields (NeRF)~\cite{nerf}. They usually employ a multi-layer perceptron (MLP) to learn the mapping from the point position to the radiodensity and then create projections by volume rendering along rays. This ray tracing scheme is very time-consuming. Even the efficient NeRF-based method NAF~\cite{naf} still requires over an hour for training and yields suboptimal results at a slow inference speed of 2 fps. This increases the waiting time of patients and doctors, leading to low diagnostic efficiency.

Recently, 3D Gaussian splatting (3DGS)~\cite{3dgs} has demonstrated promising reconstruction quality while enjoying much faster inference speed than NeRF-based algorithms in RGB domain, which motivates us to follow this technical route. However, due to the fundamental differences between X-ray and natural light imaging, directly applying the original 3DGS to X-ray NVS and direct CT reconstruction may encounter three issues. \textbf{Firstly}, the spherical harmonics (SH) in RGB 3DGS is not suitable for modeling the X-ray radiation intensity of 3D points. Specifically, natural light imaging relies on the reflection off the surface. The color of a 3D point is anisotropic and view-dependent. Based on this nature, the original Gaussian point cloud model uses SH to fit the illumination distribution.  In contrast, X-rays penetrate the object and attenuate, thereby forming an image. Given specific X-rays, the radiation intensity of a 3D point depends on its radiodensity and is independent to the view direction, which means the point radiation intensity is isotropic. \textbf{Secondly}, conventional 3DGS cannot directly reconstruct CT volumes. Its visibility-based image synthesis formulation lacks queryable 3D attenuation field required for volumetric reconstruction. \textbf{Thirdly}, the original point cloud initilization algorithm, structure-from-motion (SfM)~\cite{sfm}, is also not suitable for X-ray imaging as X-ray images are grayscale and their contrast is lower. besides, running SfM is time-consuming.

To address the above issues, we propose a novel 3DGS-based method, X-Gaussian, for X-ray NVS and direct CT reconstruction. Our X-Gaussian comprises three key techniques. \textbf{Firstly}, we redesign a radiative Gaussian point cloud model inspired by the isotropic property of X-ray imaging. We present a Radiation Intensity Response Function (RIRF) to replace the SH function of the original 3DGS. Different from SH, our RIRF excludes the influence of view direction. To this end, it adopts the inner product between a learnable vector representing the inherent point features and a set of basis weights to fit the radiation intensity of a 3D point. Based on this point cloud model, we further develop a Differentiable Radiative Rasterization (DRR) with a CUDA implementation to render novel projections. \textbf{Secondly}, we propose Radiative Gaussian Volume Querying (RGVQ), which defines a continuous attenuation field from our radiative features and opacities. It directly reconstructs CT volumes through CUDA-based parallel querying at voxel centers, without additional learnable point attributes. \textbf{Thirdly}, we customize an Angle-pose Cuboid Uniform Initialization (ACUI) strategy for camera calibration parameters and Gaussian point clouds. Our ACUI first exploits the parameters of the X-ray scanner to compute the intrinsic and extrinsic matrices. Then we set up a cuboid that can completely enclose the scanned object. Within this cuboid, we uniformly sample 3D points at intervals to initialize the center positions of the Gaussian point clouds. Free from running the  SfM algorithm, our ACUI significantly reduces the training time. Equipped with these techniques, our X-Gaussian outperforms NAF by \textbf{6.5 dB} while enjoying \textbf{73$\times$} inference speed and \textbf{7$\times$} training speed on X-ray NVS. On five synthetic scenes and three real scans, our X-Gaussian achieves \textbf{33.25 dB} and \textbf{39.53 dB} average 3D PSNR, surpassing state-of-the-art methods by \textbf{0.40 dB} and \textbf{4.34 dB}, respectively. Moreover, the CT reconstruction inference speed of X-Gaussian is \textbf{4.30$\times$} faster than SAX-NeRF on the same GPU.

The main contributions of this work can be summarized as follows:
\begin{itemize}
    \vspace{-2mm}
	\item We propose a novel 3D Gaussian splatting-based framework, X-Gaussian, for X-ray novel view synthesis. To our knowledge, this is the first attempt to explore the potential of Gaussian splatting in X-ray 3D reconstruction.
    \vspace{0.5mm}
	\item We design a radiative Gaussian point cloud model with a CUDA-based differentiable radiative rasterization for the isotropic nature of X-ray imaging.
    \vspace{-3.5mm}
	\item We formulate Radiative Gaussian Volume Querying (RGVQ) to reconstruct CT volumes from existing radiative point attributes via parallel CUDA queries, without additional learnable density parameters.
    \vspace{0.5mm}
	\item We present an angle-pose cuboid uniform initialization strategy for Gaussian point clouds and camera calibration in circular cone beam X-ray scanning.
    \vspace{0.5mm}
	\item Our X-Gaussian achieves SOTA performance in both X-ray NVS and direct CT reconstruction with substantially faster inference. It also improves projection-assisted sparse-view CT reconstruction for traditional methods.
\end{itemize}

\section{Related Work}
\vspace{-1.5mm}

\subsection{Neural Radiance Field}
\vspace{-1.5mm}

NeRF~\cite{nerf} learns an implicit neural scene representation of color and volume density, given the position of a 3D point and view direction. It has achieved great success in NVS and inspired an explosion of follow-up papers to improve its quality~\cite{mip_nerf,mipnerf360,refnerf,trimiprf,zipnerf} and speed~\cite{instant_ngp,merf,tensorf,nerfacc,bakedsdf,mobilenerf,EfficientNeRF}. Some later works extend the application domain of NeRF from natural light to X-rays~\cite{intratomo,naf,mednerf}. For instance, 
NAF~\cite{naf} follows the settings of Instant-NGP to learn the implicit mapping from 3D position to attenuation. SAX-NeRF~\cite{sax_nerf} combines line-based attention and structured ray sampling to improve sparse-view CT reconstruction. Yet, the ray tracing and volume rendering schemes are time-consuming, which limits the training and inference speed of NeRF-based X-ray 3D reconstruction. 

\vspace{-3mm}
\subsection{Gaussian Splatting}
\vspace{-1mm}
3DGS~\cite{3dgs} represents scenes using millions of 3D Gaussian point clouds. This approach is fundamentally different from NeRF-based algorithms by employing an explicit representation coupled with highly parallelized rasterization workflows. These features enable more efficient computation and rendering processes. Hence, 3DGS has achieved great success in several fields, including 3D  Generation~\cite{dreamgaussian,gaussiandreamer,luciddreamer,humangaussian,hugs,gauhuman,physdreamer}, Dynamic Scene Modeling~\cite{dynamic1,dynamic2,dynamic3}, SLAM~\cite{slam1,slam2,slam3,slam4}, Inverse Rendering~\cite{InverseRendering1,InverseRendering2,InverseRendering3}, \emph{etc.} However, most existing 3DGS algorithms are mainly designed for RGB scenes. Their view-dependent appearance conflicts with isotropic X-ray attenuation, while their visibility-based rendering lacks a queryable 3D attenuation field for direct 3D CT volume reconstruction.

\vspace{-2.5mm}
\section{Method}
\begin{figure*}[t]
	\begin{center}
		\begin{tabular}[t]{c} \hspace{-3.3mm}	\includegraphics[width=1\textwidth]{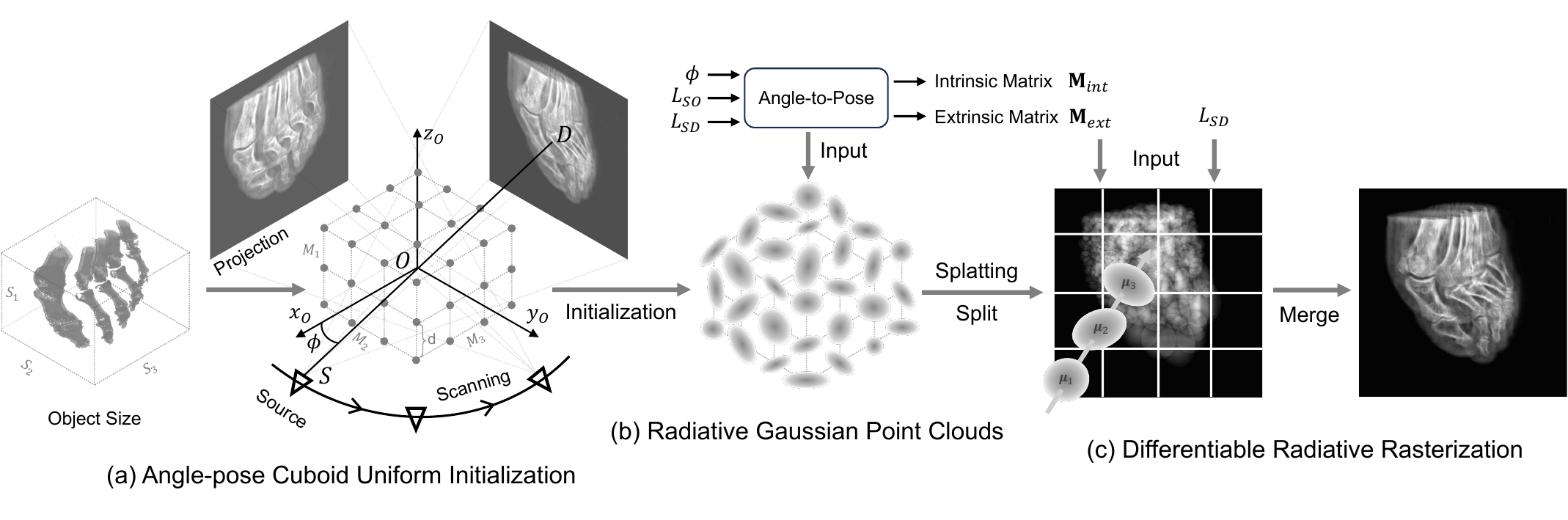}
		\end{tabular}
	\end{center}
    \vspace{-4mm}
	\caption{Pipeline of X-Gaussian. (a) Angle-pose Cuboid Uniform Initialization (ACUI) strategy uses the parameters of X-ray scanner to compute intrinsic and extrinsic matrices, and samples center points for 3D Gaussians. (b) Our radiative Gaussian point cloud model learns to predict the radiation intensity of 3D points. (c) Based on our Gaussian model, we develop a GPU-friendly Differentiable Radiative Rasterization (DRR).}
    \vspace{-1mm}
	\label{fig:pipeline}
\end{figure*}

\vspace{-2.5mm}
The pipeline of our X-Gaussian for X-ray NVS is shown in Fig.~\ref{fig:pipeline}. Firstly, we design an Angle-pose Cuboid Uniform Initialization (ACUI) to compute the intrinsic and extrinsic matrices from the parameters of X-ray scanner, as illustrated in Fig.~\ref{fig:pipeline} (a). Then ACUI uniformly samples 3D points within a cuboid that can completely enclose the scanned object to initialize the center positions of our radiative Gaussian point clouds in Fig.~\ref{fig:pipeline} (b). Given a view direction, the 3D point clouds undergo our Differentiable Radiative Rasterization (DRR) to derive the rendered image, as depicted in Fig.~\ref{fig:pipeline} (c). In this section, we first introduce our radiative Gaussian point cloud model and DRR, then describe how to directly query a CT volume from the same point attributes, and finally present ACUI.

\vspace{-2.5mm}
\subsection{Radiative Gaussian Point Cloud Model}
\label{sec:pc_model}
\vspace{-1mm}

An object can be represented by a set of basic Gaussian point clouds $\mathcal{G}$ as
{\small
\begin{equation}
    \mathcal{G} = \{G_i(\bm{\mu}_i, \mathbf{\Sigma}_i, \alpha_i)~|~i = 1, 2, \dots, N_p\},
\end{equation}
}
where $G_i$ refers to the $i$-th Gaussian point cloud. Its center position, covariance, and opacity are defined as $\bm{\mu}_i \in \mathbb{R}^3$, $\mathbf{\Sigma}_i \in \mathbb{R}^{3\times3}$, and $\alpha_i \in \mathbb{R}$. 
$\mathrm{\bf \Sigma}_i$ is represented by a rotation matrix $\mathbf{R}_i \in \mathrm{SO}(3)$ and a positive diagonal scaling matrix $\mathbf{S}_i \in \mathbb{R}^{3\times3}$ as $\mathrm{\bf \Sigma}_i = \mathbf{R}_i \mathbf{S}_i \mathbf{S}_i^\top \mathbf{R}_i^\top$. $\bm{\mu}_i$, $\mathbf{\Sigma}_i$, $\alpha_i$, $\mathbf{R}_i$, and $\mathbf{S}_i$ are learnable parameters. Besides these attributes, each Gaussian also employs additional learnable parameters to fit different imaging scenarios, \emph{e.g.}, natural light imaging and X-ray imaging.

\begin{figure*}[t]
	\begin{center}
		\begin{tabular}[t]{c} \hspace{-3.3mm}	\includegraphics[width=1.01\textwidth]{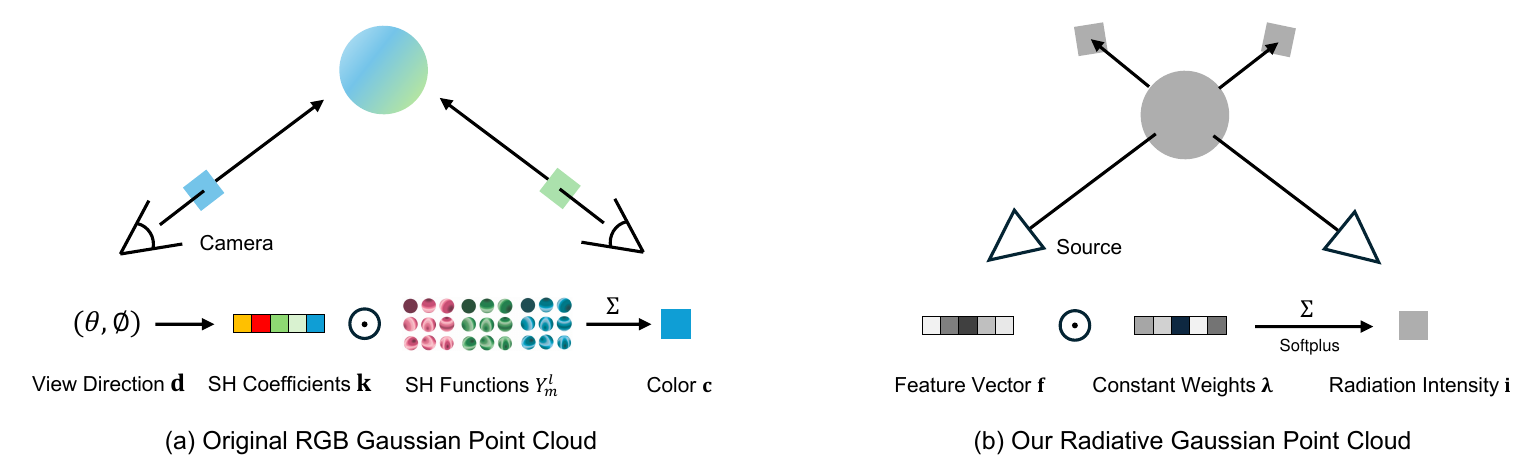}
		\end{tabular}
	\end{center}
    \vspace{-4mm}
	\caption{Comparison between the Gaussian point cloud models of the original 3DGS and our X-Gaussian. (a) The original RGB Gaussian point cloud model uses spherical harmonics (SH) to simulate the anisotropic natural light distribution and view-dependent color. (b) Our radiative Gaussian point cloud model employs the weighted sum of point features to fit the isotropic X-ray penetration and view-independent radiation intensity.}
    \vspace{-2mm}
	\label{fig:pc_model}
\end{figure*}

We first review the original RGB Gaussian point cloud model~\cite{3dgs} in natural light imaging.  As shown in Fig.~\ref{fig:pc_model} (a),  the color of a 3D point is represented by spherical harmonics (SH). The point color is anisotropic and changes with the view direction. Each Gaussian point cloud learns to predict the SH coefficients $\mathbf{k} = \{k_l^m | 0 \leq l \leq L, -l \leq m \leq l\} \in \mathbb{R}^{(L+1)^2\times 3}$, where each $k_l^m \in \mathbb{R}^{3}$ is a set of 3 coefficients corresponding to the RGB components. $L$ is the degree of SH. Then the point color $\mathbf{c} \in \mathbb{R}^{3}$ at the view direction $\mathbf{d} = (\theta, \phi)$ is derived by
\vspace{-1.5mm}
{\small
\begin{equation}
	\mathbf{c}(\mathbf{d}, \mathbf{k}) = \sum_{l  = 0}^{L} \sum_{m = -l}^{l} k_l^m~Y_l^m(\theta, \phi),
	\label{eq:original_pc_model}
\vspace{-1.5mm}
\end{equation}
}
where  $Y_l^m: \mathbb{S}^2 \rightarrow \mathbb{R}$ is the SH function that maps points on the sphere to real numbers. Please refer to the supplementary for its detailed formulation.


Although 3DGS~\cite{3dgs} achieves fast inference speed and good performance in natural light imaging, the RGB Gaussian point cloud model is not suitable for X-ray scenarios due to the fundamental differences between natural light imaging and X-ray imaging. Natural light imaging relies on the reflection off the surface of object. The anisotropic color modeled by SH is view-dependent. \emph{e.g.}, in Fig.~\ref{fig:pc_model} (a), the point color is blue from the left viewpoint and green from the right viewpoint. In contrast, X-ray imaging is based on the attenuation when penetrating the object. The degree of attenuation depends on the isotropic radiodensity property. Thus, the radiation intensity of a 3D point is view-independent. 

Based on this analysis, we redesign our radiative Gaussian point cloud model by replacing the original 3DGS's SH color representation with a Radiation Intensity Response Function (RIRF) to predict point radiation intensity. As shown in Fig.~\ref{fig:pc_model} (b), each Gaussian learns radiative features $\mathbf{f}_i \in \mathbb{R}^{N_f\times3}$, with $N_f$ coefficients per channel, to encode its inherent properties. Define its view-independent scalar radiation response and opacity-weighted attenuation coefficient as
{\small
\begin{equation}
        \bar{\mathbf{i}}_i = \mathrm{RIRF}(\mathbf{f}_i)
        = \frac{1}{3}\sum_{c=1}^{3}\psi\!\left(\frac{1}{2}+\sum_{j=1}^{N_f}\lambda_j f_{ijc}\right),
        ~~~~\rho_i = \alpha_i\bar{\mathbf{i}}_i,
	\label{eq:xray_pc_model}
\end{equation}
}
where $f_{ijc}$ denotes the $j$-th feature of the $c$-th stored channel of point $i$, and $\bm{\lambda} \in \mathbb{R}^{N_f}$ contains constant, view-independent basis weights. The Softplus function $\psi(z)=\log(1+\exp(z))$ ensures a positive response, and $\alpha_i$ is the sigmoid-activated opacity. The attenuation coefficient $\rho_i$ is computed from these existing attributes rather than introduced as an additional learnable parameter. Then the set of our radiative Gaussian point clouds $\mathcal{G}_x$ is formulated as
{\small
\begin{equation}
    \mathcal{G}_x = \{G_i(\bm{\mu}_i, \mathbf{\Sigma}_i, \alpha_i, \mathbf{f}_i)~|~i = 1, 2, \dots, N_p\},
\end{equation}
}
where $\mathbf{f}_i \in \mathbb{R}^{N_f\times3}$ denotes the radiative features of the $i$-th Gaussian point cloud. Please note that Eq.~\eqref{eq:xray_pc_model} excludes the influence of the view direction $\mathbf{d} = (\theta, \phi)$, which matches the isotropic nature of X-ray imaging. Meanwhile, Eq.~\eqref{eq:xray_pc_model} is free from the complex computation of SH function.
Hence, the forward and backward processes of our X-Gaussian are much faster than those of the original 3DGS. 

\vspace{-2mm}
\subsection{Differentiable Radiative Rasterization}
\vspace{-1mm}
Based on our radiative Gaussian point cloud, we develop a Differentiable Radiative Rasterization (DRR), as shown in Fig.~\ref{fig:pipeline} (c). We first summarize the overall DRR processing $F_{\text{DRR}}$ and then describe its details. DRR is represented as
{\small
\begin{equation}
\mathbf{I} = F_{\text{DRR}}(\mathbf{M}_{ext}, \mathbf{M}_{int}, \{G_i(\bm{\mu}_i, \mathbf{\Sigma}_i, \alpha_i, \mathbf{f}_i)~|~{i = 1, 2, \dots, N_p}\}),
 \label{eq:drr}
\end{equation}
}
where $\mathbf{I} \in \mathbb{R}^{H\times W}$ denotes the rendered image, $\mathbf{M}_{ext} \in \mathbb{R}^{4\times 4}$ represents the extrinsic matrix, and $\mathbf{M}_{int} \in \mathbb{R}^{3\times 4}$ refers to the intrinsic matrix. Subsequently, we introduce the details of $F_{\text{DRR}}$. To begin with, the possibility value of the $i$-th Gaussian distribution at the 3D point position $\mathbf{x} \in \mathbb{R}^3$ is formulated as
{\small
\begin{equation}
	P(\mathbf{x}|\mathrm{\bm{\mu}}_i, \mathrm{\bf \Sigma}_i) = \text{exp}\big(-\frac{1}{2}(\mathbf{x} - \bm{\mu}_i)^\top \mathrm{\bf \Sigma}_i^{-1} (\mathbf{x} - \bm{\mu}_i)\big).
\end{equation}
}
Then we project the 3D Gaussians to the 2D detector plane for subsequent rendering.  $\bm{\mu}_i$ is firstly transferred from the world coordinate system to the camera coordinate system and then projected to the image coordinate system as
{\small
\begin{equation}
	\widetilde{\mathbf{t}}_i = 
	\begin{bmatrix}
		\mathbf{t}_{i}\\
		1
	\end{bmatrix} = \mathbf{M}_{ext}~ \widetilde{\bm{\mu}}_i = \mathbf{M}_{ext}~
	\begin{bmatrix}
		\bm{\mu}_{i}\\
		1
	\end{bmatrix}, ~~~~
    \widetilde{\mathbf{u}}_i = 
	\begin{bmatrix}
		\mathbf{u}_{i}\\
		1
	\end{bmatrix} = \frac{1}{t_z}\mathbf{M}_{int}~\widetilde{\mathbf{t}}_i,
\end{equation}
}
where $\mathbf{t}_{i} = ({t}_{x}, {t}_{y}, {t}_{z}) \in \mathbb{R}^3$ is the camera coordinate of $\bm{\mu}_{i}$ with $t_z>0$, and $\mathbf{u}_i \in \mathbb{R}^{2}$ is its image coordinate in pixels. $\widetilde{\mathbf{u}}_i$, $\widetilde{\mathbf{t}}_i$, and $\widetilde{\bm{\mu}}_i$ are their homogeneous coordinates; the factor $1/t_z$ performs perspective division so that the last component of $\widetilde{\mathbf{u}}_i$ is one. Subsequently, we transfer the 3D covariance matrix $\mathrm{\bf \Sigma}_i$ to its counterpart $\mathrm{\bf \Sigma}_i^{'} \in \mathbb{R}^{3\times 3}$ in the projected coordinate system as
{\small
\begin{equation}
	\mathrm{\bf \Sigma}_i^{'} = \mathbf{J}_i \mathbf{W}_i \mathrm{\bf \Sigma}_i \mathbf{W}_i^\top \mathbf{J}_i^\top,
\end{equation}
}
where $\mathbf{J}_i \in \mathbb{R}^{3\times 3}$ is the Jacobian of the affine approximation of the projective transformation. $\mathbf{W}_i \in \mathbb{R}^{3\times 3}$ is the viewing transformation. We derive them by
{\small
\begin{equation}
	\mathbf{J}_i = 
	\begin{bmatrix}
		\frac{f_x}{t_z} &0 &-\frac{f_x t_x}{t_z^2}\\
		0 & \frac{f_y}{t_z} &-\frac{f_y t_y}{t_z^2} \\
		0 & 0 & 0
	\end{bmatrix},~~~~~~
	\mathbf{W}_{i} =
	\begin{bmatrix}
		-\sin\phi & \cos\phi & 0 \\
		0 & 0 & -1 \\
		-\cos\phi & -\sin\phi & 0  \\
	\end{bmatrix},
\end{equation}
}
where $f_x=L_{SD}/\Delta_x$ and $f_y=L_{SD}/\Delta_y$ are the focal lengths in pixels; the source-to-detector distance $L_{SD}$ and detector pixel spacings $\Delta_x,\Delta_y$ use the same physical length units. $\phi$ is the azimuth angle of the source. Following ~\cite{ewa,kopanas2021point,3dgs}, we obtain the 2D covariance matrix $\mathrm{\bf \Sigma}_i^{''} \in \mathbb{R}^{2\times2}$ by skipping the third row and column of $\mathrm{\bf \Sigma}_i^{'}$. Then the 2D projection is partitioned into non-overlapping tiles. The 3D Gaussians ($\bm{\mu}_i$,$\bm{\mathbf{\Sigma}}_i$) are assigned to different tiles according to their 2D projections ($\bm{u}_i$,$\bm{\mathbf{\Sigma}}^{''}_i$), as shown in the left image of Fig.~\ref{fig:pipeline} (c). Under the local affine projection model, DRR computes the attenuation projection at pixel $p$ by accumulating the integrated Gaussian contributions as
{\small
\begin{equation}
    \mathbf{I}(p)=\sum_{i\in\mathcal{N}(p)}\rho_i\ell_i\exp\!\left[-\tfrac{1}{2}(p-\mathbf{u}_i)^\top(\mathbf{\Sigma}_i^{''})^{-1}(p-\mathbf{u}_i)\right],\quad \ell_i=\sqrt{\tfrac{2\pi}{\mathbf{d}_i^\top\mathbf{\Sigma}_i^{-1}\mathbf{d}_i}},
    \label{eq:drr_projection}
\end{equation}
}
where $\mathcal{N}(p)$ contains the Gaussians contributing to pixel $p\in\mathbb{R}^2$, and $\rho_i=\alpha_i\bar{\mathbf{i}}_i$ is the attenuation amplitude of Gaussian $i$, computed from its opacity and RIRF response in Eq.~\eqref{eq:xray_pc_model}. The unit vector $\mathbf{d}_i$ points from the X-ray source to $\bm{\mu}_i$ in world coordinates, and $\ell_i$ accounts for Gaussian integration along this direction. This local projection model approximates ray integrals of the continuous attenuation field defined in Sec.~\ref{sec:direct_ct_method}.

\noindent\textbf{Optimization.} Eventually, the  training objective $\mathcal{L}$ is the weighted sum of $\mathcal{L}_1$ loss and SSIM loss between the rendered and ground-truth projection images as
\vspace{-1mm}
{\small
\begin{equation}
	\mathcal{L} = (1 - \gamma) \mathcal{L}_1 + \gamma \mathcal{L}_{\text{SSIM}},
\label{eq:loss}
\vspace{1mm}
\end{equation}
}
where $\gamma$ is a hyperparameter balancing the importances of the two loss terms. By minimizing Eq.~\eqref{eq:loss}, we can optimize the attributes of 3D Gaussians, \emph{i.e.}, $\bm{\mu}_i, \mathbf{\Sigma}_i, \alpha_i, \text{and} ~\mathbf{f}_i$ in Eq.~\eqref{eq:drr}. $N_p$ is adjusted by the adaptive control~\cite{3dgs}.

Compared to the RGB rasterization in 3DGS~\cite{3dgs}, our DRR avoids the complex computations related to the view direction in the forward and backward processes, thereby enjoying cheaper training costs and faster inference speed.

\subsection{Radiative Gaussian Volume Querying}
\label{sec:direct_ct_method}
Conventional 3DGS cannot directly reconstruct CT volumes because its opacities control visibility and depth-ordered image blending. These opacities do not directly represent the physical attenuation coefficients needed to define a consistent 3D field for CT reconstruction. To address this limitation, we propose Radiative Gaussian Volume Querying (RGVQ) to directly reconstruct CT volumes from our radiative point clouds. It retains the original point positions, covariances, opacities, and radiative features, and uses the attenuation coefficient $\rho_i$ defined by RIRF in Eq.~\eqref{eq:xray_pc_model}. Since the Gaussian function $P(\mathbf{x}|\bm{\mu}_i,\mathbf{\Sigma}_i)$ has unit value at its center, the contribution of point $i$ is $\rho_i P(\mathbf{x}|\bm{\mu}_i,\mathbf{\Sigma}_i)$. Summing the contributions gives a continuous attenuation field as
{\small
\begin{equation}
    V(\mathbf{x}) = \sum_{i=1}^{N_p}\rho_i
    \exp\!\left[-\frac{1}{2}(\mathbf{x}-\bm{\mu}_i)^\top
    \mathbf{\Sigma}_i^{-1}(\mathbf{x}-\bm{\mu}_i)\right],
    \label{eq:ct_field}
\end{equation}
}
where $\mathbf{\Sigma}_i=\mathbf{R}_i\mathbf{S}_i\mathbf{S}_i^\top\mathbf{R}_i^\top$ reuses the original rotations and scales. Unlike surface visibility-based blending, Eq.~\eqref{eq:ct_field} adds attenuation contributions throughout the volume and is independent of the querying view direction. We discretize this field by sampling voxel centers. For a reconstruction grid with center $\mathbf{c}$, physical size $\mathbf{S}$, and resolution $\mathbf{M}=(M_1,M_2,M_3)$, voxel $\mathbf{k}$ is reconstructed as
{\small
\begin{equation}
    \mathbf{x}_{\mathbf{k}}=\mathbf{c}-\frac{\mathbf{S}}{2}
    +\left(\mathbf{k}+\frac{\mathbf{1}}{2}\right)\odot\frac{\mathbf{S}}{\mathbf{M}},
    \qquad \mathbf{V}[\mathbf{k}]=V(\mathbf{x}_{\mathbf{k}}),
    \label{eq:ct_query}
\end{equation}
}
where $0\leq k_d<M_d$ and the multiplication and division are element-wise. We implement our RGVQ algorithm through CUDA-based parallel accumulation of Gaussian contributions at voxel centers. This efficient query remains differentiable with respect to the original Gaussian point cloud attributes and shares the attenuation response in Eq.~\eqref{eq:xray_pc_model} with the 2D projection rendering.

\subsection{Angle-pose Cuboid Uniform Initialization}
We need to initialize the parameters in Eq.~\eqref{eq:drr} for rasterization. Specifically, $\mathbf{\Sigma}_i$, $\alpha_i$, and $\mathbf{f}_i$ are randomly initialized. In natural light imaging, the original 3DGS~\cite{3dgs} adopts the SfM~\cite{sfm} algorithm to compute the initial $\bm{\mu}_i$, $N_p$, $\mathbf{M}_{ext}$, and $\mathbf{M}_{int}$. SfM detects and matches features from multi-view images. It is not suitable for X-ray imaging due to two reasons. Firstly, X-ray images are grayscale and low-contrast. Secondly, different layers of an object may overlap on the same positions of the projection. These two problems degrade the accuracy of feature detection and matching in SfM. Besides, running the SfM algorithm usually requires a long time, which prolongs the training process.

To address these issues, we customize an Angle-pose Cuboid Uniform Initialization (ACUI) strategy for circular cone beam X-ray scanning scenario where a scanner emits cone-shaped X-ray beams and captures projections at equal angular intervals. As shown in Fig.~\ref{fig:pipeline} (a), ACUI uses the parameters of X-ray scanner  to compute the extrinsic matrix $\mathbf{M}_{ext}$ and intrinsic matrix $\mathbf{M}_{int}$ as
{\small
\begin{equation}
        \mathbf{M}_{ext} = 
        \begin{bmatrix}
            -\sin\phi & \cos\phi & 0 & 0 \\
            0 & 0 & -1 & 0 \\
            -\cos\phi & -\sin\phi & 0 & L_{SO} \\
            0 & 0 & 0 & 1
        \end{bmatrix}, ~~~~
        \mathbf{M}_{int} = 
        \begin{bmatrix}
            f_x & 0 & {W} / {2} &~~0~~ \\
            0 & f_y & {H} / {2} &~~0~~ \\
            0 & 0 & 1 &~~0~~
        \end{bmatrix},
\end{equation}
}
where $L_{SO}$ represents the distance between the X-ray source and the scanned object, and $f_x,f_y$ are the pixel focal lengths defined above. The elevation angle of the X-ray source is set to zero and remains unchanged.
The next step of ACUI is to initialize the center positions of 3D Gaussians. Although the precise shape of the scanned object is not given at the beginning, the scanning space can be approximated. We set up a cuboid with size $S_1\times S_2 \times S_3$ (mm)  that can completely enclose the object. The center of this cuboid is also the center of the object and the origin of the world coordinate system. We divide this cuboid by a grid with size $M_1\times M_2\times M_3$ (voxel). Then we uniformly sample points as
{\small
\begin{equation}
	\mathcal{P} = \Big\{\big(\frac{n_1 S_1 d}{M_1}, \frac{n_2 S_2 d}{M_2}, \frac{n_3 S_3 d}{M_3}\big)~\big|~-\big[\frac{M_i}{2d}\big]-1 \leq n_i \leq \big[\frac{M_i}{2d}\big]+1, ~i = 1, 2, 3\Big\},
	\label{eq:sampling}
\end{equation}
}
where $n_i \in \mathbb{Z}$. Then we use the size and elements of $\mathcal{P}$ to initialize $N_p$ and $\bm{\mu}_i$. Avoiding running SfM, ACUI allows X-Gaussian to enjoy a faster training speed.

\begin{table*}[t]
	\renewcommand{\arraystretch}{1.0}
	\newcommand{\tabincell}[2]{\begin{tabular}{@{}#1@{}}#2\end{tabular}}
	\centering
	\renewcommand{\arraystretch}{1.55}
        \caption{Quantitative results on the novel view synthesis task. The average inference speed and training time of all scenes evaluated on an RTX 8000 GPU are reported. In the cell of the results of each scene, PSNR (upper) and SSIM (lower) are listed.}
	\resizebox{\textwidth}{!}
	{
		\centering
		\begin{tabular}{lccccccccccccc}
			\toprule[0.2em]
			\rowcolor{lightgray}
			Method~~ & ~Infer Speed~ & ~Train Time~ & ~Chest~ & ~Foot~ & ~Head~ & ~Abdomen~ &~Pancreas~ & ~Average~ \\ \midrule[0.1em]
			InTomo~\cite{intratomo} & 0.62 fps & 125 min & \tabincell{c}{28.948\\0.9915} & \tabincell{c}{39.482\\0.9979} &\tabincell{c}{34.832\\0.9977}  & \tabincell{c}{27.641\\0.9646} &\tabincell{c}{20.031\\0.8537} &\tabincell{c}{30.187\\0.9611}  \\
			\midrule[0.1em]
			NeRF~\cite{nerf} & 0.14 fps & 313 min & \tabincell{c}{36.157\\0.9988} & \tabincell{c}{{41.053}\\{0.9989}} & \tabincell{c}{29.760\\0.9991}
			& \tabincell{c}{24.620\\0.9559} &\tabincell{c}{19.853\\0.8560} &\tabincell{c}{30.289\\0.9617} \\
			\midrule[0.1em]
			TensoRF~\cite{tensorf}  & 0.77 fps & 178 min & \tabincell{c}{23.609\\0.9402} & \tabincell{c}{37.728\\0.9929} &\tabincell{c}{34.429\\0.9879}  & \tabincell{c}{27.382\\0.8730} &\tabincell{c}{29.235\\0.8031}  &\tabincell{c}{30.477\\0.9194} \\
			\midrule[0.1em]
			NeAT~\cite{neat} & 1.78 fps & 69 min & \tabincell{c}{40.765\\0.9990} & \tabincell{c}{38.236\\0.9963} &\tabincell{c}{27.738\\0.9295}  & \tabincell{c}{26.741\\0.8563} &\tabincell{c}{37.526\\0.9017} &\tabincell{c}{34.201\\0.9366} \\
			\midrule[0.1em]
			NAF~\cite{naf}  & {2.01 fps} & {63 min} & \tabincell{c}{{42.366}\\{0.9993}} & \tabincell{c}{38.353\\0.9913} & \tabincell{c}{30.174\\0.9531} &\tabincell{c}{{37.590}\\{0.9855}} &\tabincell{c}{36.228\\0.8844} &\tabincell{c}{36.942\\0.9627} \\
			\midrule[0.1em]
			\rowcolor{rouse}
			X-Gaussian & \bf 148 fps & \bf 9 min & \tabincell{c}{\bf 43.887\\ \bf 0.9998} & \tabincell{c}{\bf 42.153 \\ \bf 0.9997} & \tabincell{c}{\bf 41.579 \\ \bf 0.9997} & \tabincell{c}{\bf 45.762 \\ \bf 0.9999} & \tabincell{c}{\bf 43.640 \\ \bf 0.9976} & \tabincell{c}{\bf 43.404 \\ \bf 0.9993}\\
			\bottomrule[0.2em]
		\end{tabular}
	}
	\label{tab:proj_compare}
\end{table*}
\vspace{-3mm}

\section{Experiments}

\subsection{Experimental Settings}

\subsubsection{Dataset.} Following NAF~\cite{naf}, we use the public datasets of human organ CTs, \emph{i.e.}, LIDC-IDRI~\cite{LIDC-IDRI} and the open scientific visualization dataset~\cite{osvd}, to evaluate our method on synthetic scenes. The test scenes include chest, foot, head, abdomen, and pancreas. For these synthetic scenes, we use the open-source tomographic toolbox TIGRE~\cite{tigre} to capture 100 projections with 3\% noise in the range of 0 $\sim$ 180$^\circ$. 50 projections are used for training and the other 50 projections are used for testing. We further evaluate three real cone-beam CT scans from the Finnish Inverse Problems Society, namely Pine, Seashell, and Walnut~\cite{fips_pine,fips_seashell,fips_walnut}. Each scene uses 50 training and 100 test projections at $560\times560$ resolution, with a $256^3$ reconstruction grid. We use the full-view FDK~\cite{fdk} CT reconstructions as the ground truth for quantitative evaluation.

\begin{figure*}[t]
	\begin{center}
		\begin{tabular}[t]{c} \hspace{-1.6mm}
	\includegraphics[width=1\textwidth]{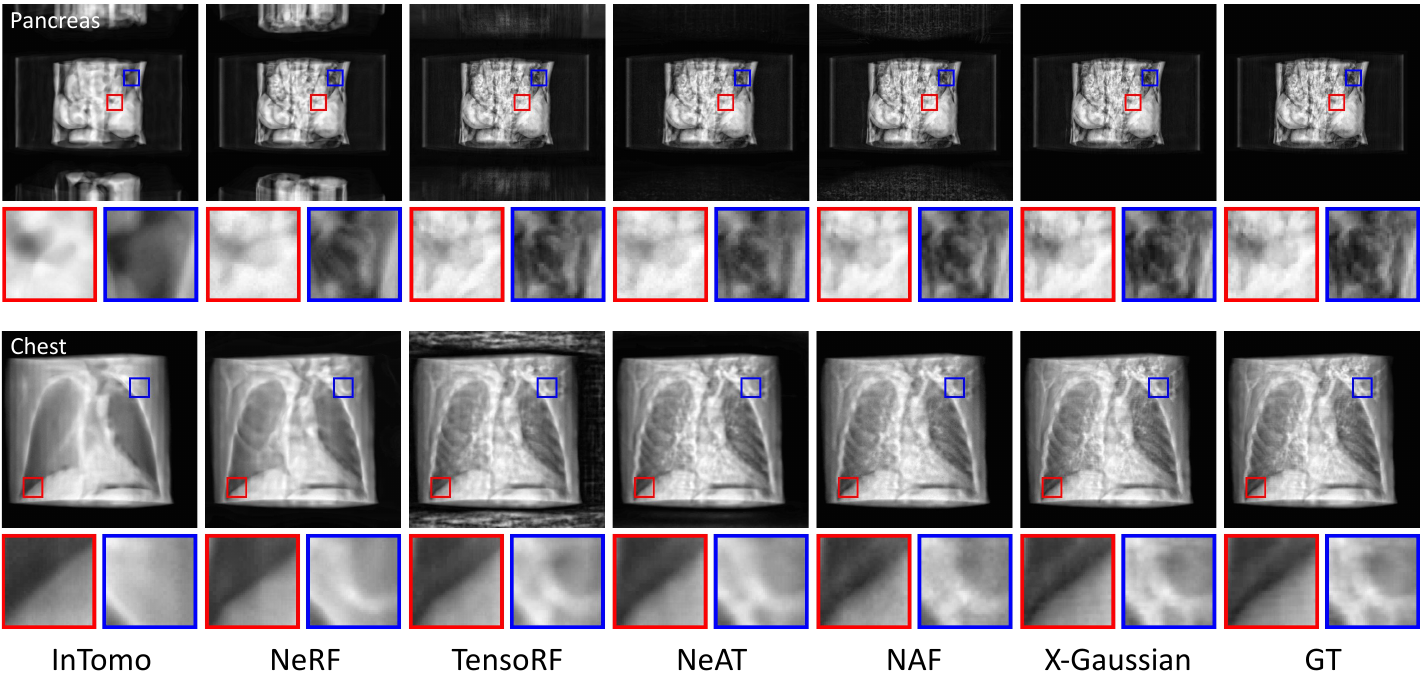}
		\end{tabular}
	\end{center}
    \vspace{-3mm}
	\caption{Qualitative results of novel view synthesis on the scenes of pancreas (top) and chest (bottom). Our X-Gaussian yields clearer results. Please zoom in for a better view.}
	\label{fig:proj_compare_1}
\end{figure*}

\subsubsection{Implementation Details.} X-Gaussian is implemented by PyTorch~\cite{pytorch} and CUDA~\cite{cuda}. The model is trained with the Adam optimizer~\cite{adam} ($\beta_1$ = 0.9, $\beta_2$ = 0.999, and $\epsilon$ = 1$\times$10$^{-15}$) for 2$\times$10$^{4}$ iterations. The learning rate for point cloud position is initially set to 1.9$\times$10$^{-4}$ and exponentially decays to 1.9$\times$10$^{-6}$. The learning rates for point feature, opacity, scaling, and rotation are set to 2$\times$10$^{-3}$, 8$\times$10$^{-3}$, 5$\times$10$^{-3}$, and 1$\times$10$^{-3}$.   $\gamma$ in Eq.~\eqref{eq:loss} is set to 0.2. We adopt peak signal-to-noise ratio (PSNR) and structural similarity index measure (SSIM)~\cite{ssim} to evaluate the performance.  Frames per second (fps) is used to measure the inference speed. The experiments are done RTX 8000.

\subsection{Novel View Synthesis}
\subsubsection{Quantitative Results.} Tab.~\ref{tab:proj_compare} shows the quantitative comparisons between our X-Gaussian and five NeRF-based algorithms, including InTomo~\cite{intratomo}, NeRF~\cite{nerf}, TensoRF~\cite{tensorf}, NeAT~\cite{neat}, and NAF~\cite{naf} on the NVS task.

We report the average inference speed and training time of different methods on all scenes. In the cell of the results of each scene, PSNR (upper entry in the cell) and SSIM (lower entry in the cell) are listed.  As can be observed that our X-Gaussian not only surpasses SOTA methods by large margins in performance but also enjoys much faster inference speed and cheaper training costs. More specifically, compared with the efficient X-ray NeRF-based method NAF,  our X-Gaussian outperforms it by 6.5 dB on average and is 73$\times$ faster in inference while requiring less than 15\% training time. When compared with the SOTA RGB NeRF-based method TensoRF, our X-Gaussian is 12.93 dB higher while enjoying 192$\times$ inference speed and 20$\times$ training speed.

\subsubsection{Qualitative Results.} 
Figs.~\ref{fig:proj_compare_1} depicts the qualitative comparisons of NVS on the scenes of pancreas and chest. It can be observed from the zoomed-in patches that previous NeRF-based algorithms fail to render clear novel views. They either introduce undesired artifacts or produce blurry textures such as the toe bones of the foot. In contrast, our method yields visually realistic images by rendering more fine-grained details and clearer structural contents.

\begin{table*}[t]
\centering
\renewcommand{\arraystretch}{1.6}
\caption{Direct CT reconstruction from 50 projections on synthetic and real scenes. Cells report PSNR (dB, upper) and SSIM (lower); best and second-best results are bold and underlined. Real scans are evaluated against full-view FDK pseudo-references.}
\label{tab:direct_ct}
\setlength{\tabcolsep}{2.3pt}
\resizebox{\textwidth}{!}{
\begin{tabular}{lcccccccccc}
\toprule[0.2em]
\rowcolor{lightgray}
 & \multicolumn{6}{c}{Synthetic Scenes} & \multicolumn{4}{c}{Real Scenes} \\
\rowcolor{lightgray}
Method ~~~~~~~~& ~Chest~ & ~Foot~ & ~Head~ & ~Abdomen~ & ~Pancreas~ & ~Average~ & ~Pine~ & ~Seashell~ & ~Walnut~ & ~Average~ \\
\midrule[0.1em]
NeRF~\cite{nerf} & \begin{tabular}{@{}c@{}}33.911\\0.9266\end{tabular} & \begin{tabular}{@{}c@{}}30.206\\0.8387\end{tabular} & \begin{tabular}{@{}c@{}}31.760\\0.9007\end{tabular} & \begin{tabular}{@{}c@{}}30.435\\0.8326\end{tabular} & \begin{tabular}{@{}c@{}}18.808\\0.8471\end{tabular} & \begin{tabular}{@{}c@{}}29.024\\0.8691\end{tabular} & \begin{tabular}{@{}c@{}}36.766\\0.8547\end{tabular} & \begin{tabular}{@{}c@{}}36.225\\0.7999\end{tabular} & \begin{tabular}{@{}c@{}}30.779\\0.5133\end{tabular} & \begin{tabular}{@{}c@{}}34.590\\0.7226\end{tabular} \\
\midrule[0.1em]
NAF~\cite{naf} & \begin{tabular}{@{}c@{}}32.592\\0.9202\end{tabular} & \begin{tabular}{@{}c@{}}31.774\\0.8782\end{tabular} & \begin{tabular}{@{}c@{}}36.994\\0.9426\end{tabular} & \begin{tabular}{@{}c@{}}\ul{34.371}\\0.8836\end{tabular} & \begin{tabular}{@{}c@{}}19.434\\0.6599\end{tabular} & \begin{tabular}{@{}c@{}}31.033\\0.8569\end{tabular} & \begin{tabular}{@{}c@{}}40.459\\0.9264\end{tabular} & \begin{tabular}{@{}c@{}}40.521\\0.9275\end{tabular} & \begin{tabular}{@{}c@{}}\ul{32.440}\\0.6584\end{tabular} & \begin{tabular}{@{}c@{}}37.807\\0.8374\end{tabular} \\
\midrule[0.1em]
SAX-NeRF~\cite{sax_nerf} & \begin{tabular}{@{}c@{}}\ul{34.359}\\0.9378\end{tabular} & \begin{tabular}{@{}c@{}}\textbf{32.246}\\\textbf{0.8883}\end{tabular} & \begin{tabular}{@{}c@{}}39.699\\0.9714\end{tabular} & \begin{tabular}{@{}c@{}}\textbf{34.995}\\\textbf{0.9004}\end{tabular} & \begin{tabular}{@{}c@{}}\textbf{22.949}\\\textbf{0.9349}\end{tabular} & \begin{tabular}{@{}c@{}}\ul{32.850}\\\textbf{0.9266}\end{tabular} & \begin{tabular}{@{}c@{}}37.303\\0.9168\end{tabular} & \begin{tabular}{@{}c@{}}37.018\\0.9321\end{tabular} & \begin{tabular}{@{}c@{}}31.245\\0.6701\end{tabular} & \begin{tabular}{@{}c@{}}35.188\\0.8396\end{tabular} \\
\midrule[0.1em]
R$^2$-Gaussian & \begin{tabular}{@{}c@{}}33.601\\\textbf{0.9427}\end{tabular} & \begin{tabular}{@{}c@{}}32.022\\0.8799\end{tabular} & \begin{tabular}{@{}c@{}}\ul{39.944}\\\ul{0.9763}\end{tabular} & \begin{tabular}{@{}c@{}}33.498\\0.8931\end{tabular} & \begin{tabular}{@{}c@{}}21.808\\0.8577\end{tabular} & \begin{tabular}{@{}c@{}}32.175\\0.9099\end{tabular} & \begin{tabular}{@{}c@{}}\ul{41.370}\\\ul{0.9454}\end{tabular} & \begin{tabular}{@{}c@{}}\ul{43.426}\\\ul{0.9616}\end{tabular} & \begin{tabular}{@{}c@{}}32.174\\\ul{0.7082}\end{tabular} & \begin{tabular}{@{}c@{}}\ul{38.990}\\\ul{0.8717}\end{tabular} \\
\midrule[0.1em]
\rowcolor{rouse}
X-Gaussian & \begin{tabular}{@{}c@{}}\textbf{36.001}\\\ul{0.9391}\end{tabular} & \begin{tabular}{@{}c@{}}\ul{32.181}\\\ul{0.8815}\end{tabular} & \begin{tabular}{@{}c@{}}\textbf{41.438}\\\textbf{0.9768}\end{tabular} & \begin{tabular}{@{}c@{}}34.270\\\ul{0.8946}\end{tabular} & \begin{tabular}{@{}c@{}}\ul{22.378}\\\ul{0.8791}\end{tabular} & \begin{tabular}{@{}c@{}}\textbf{33.254}\\\ul{0.9142}\end{tabular} & \begin{tabular}{@{}c@{}}\textbf{41.798}\\\textbf{0.9465}\end{tabular} & \begin{tabular}{@{}c@{}}\textbf{44.008}\\\textbf{0.9638}\end{tabular} & \begin{tabular}{@{}c@{}}\textbf{32.769}\\\textbf{0.7169}\end{tabular} & \begin{tabular}{@{}c@{}}\textbf{39.525}\\\textbf{0.8757}\end{tabular} \\
\bottomrule[0.2em]
\end{tabular}}
\end{table*}

\FloatBarrier
\subsection{3D CT Reconstruction}
\label{sec:direct_ct_experiment}
\begin{figure*}[t]
    \centering
    \includegraphics[width=\textwidth]{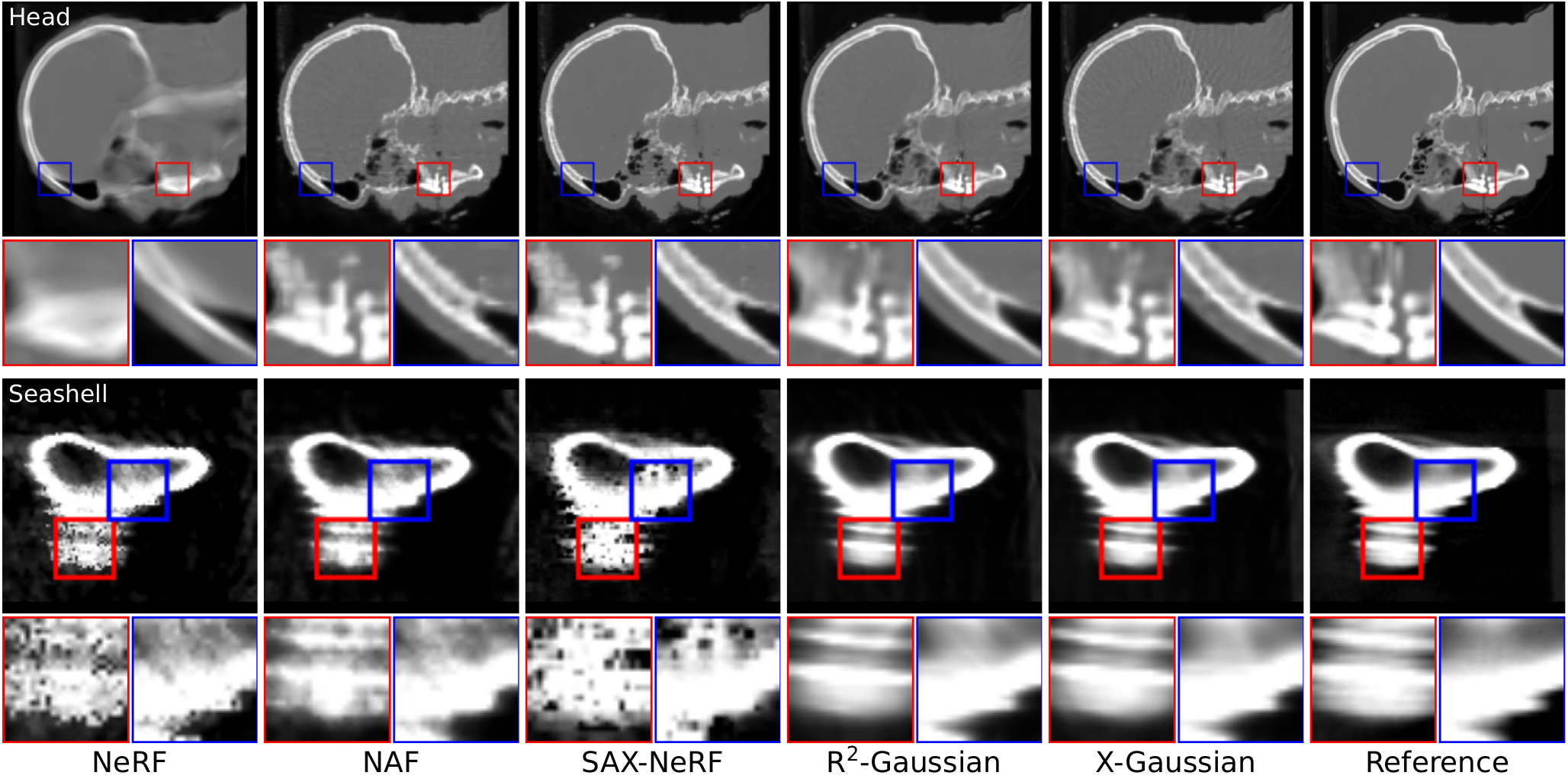}
    \caption{Visual comparisons of CT reconstruction on Head (top) and Seashell (bottom). Our method can reconstruct more structural details. Zoom in for a better view.}
    \label{fig:direct_ct}
\end{figure*}

We compare our X-Gaussian with NeRF~\cite{nerf}, NAF~\cite{naf}, SAX-NeRF~\cite{sax_nerf}, and R$^2$-Gaussian~\cite{r2_gaussian} on direct 3D CT reconstruction. Tab.~\ref{tab:direct_ct} reports the results on the five synthetic scenes. Our X-Gaussian achieves \textbf{33.25 dB} average 3D PSNR, outperforming SAX-NeRF and R$^2$-Gaussian by \textbf{0.40 dB} and \textbf{1.08 dB}, respectively. The largest gains over SAX-NeRF are obtained on Chest and Head. SAX-NeRF retains higher average SSIM on the synthetic scenes, particularly on Pancreas. Fig.~\ref{fig:direct_ct} (top) shows the CT slice comparisons on Head. The enlarged regions reveal differences in fine bone boundaries and spinal structures, while residual artifacts remain in the limited-angle reconstructions.

We further compare our X-Gaussian with SOTA methods on real scenes. We use the CT volume reconstructed by FDK as the reference ground truth for evaluation. The results are shown in Tab.~\ref{tab:direct_ct}, our X-Gaussian achieves \textbf{39.53 dB} average 3D PSNR, surpassing SAX-NeRF by \textbf{4.34 dB} and R$^2$-Gaussian by \textbf{0.54 dB}. Our method also obtains the highest SSIM on all three real scenes. Fig.~\ref{fig:direct_ct} (bottom) presents the visual comparisons on Seashell against the common FDK reference. The enlarged regions highlight local intensity transitions and shell boundaries. X-Gaussian can reconstruct more structural details.

\subsection{Projection-Assisted CT Reconstruction}

We compare our method with SOTA NeRF-based algorithms on projection-assisted sparse-view CT reconstruction. In addition to the direct volume query in Sec.~\ref{sec:direct_ct_method}, we evaluate a complementary application of X-ray NVS: using different NeRF-based methods and our X-Gaussian to create novel-view projections for three learning-free CT reconstruction methods, including an analytical method (FDK~\cite{fdk}) and two iterative methods (SART~\cite{sart} and ASD-POCS~\cite{asd_pocs}). Specifically, these three methods reconstruct the CTs from 5 original projections and 95 novel-view projections rendered by different NVS algorithms. The quantitative results are listed in Tab.~\ref{tab:ct}. When only using 5 original views (+ None), FDK, SART, and ASD-POCS achieve 7.41, 17.24, and 17.03 dB in PSNR, respectively. They fail to reconstruct the CT volumes. When employing our X-Gaussin to create novel X-ray projections for FDK, SART, and ASD-POCS, they yield the most significant improvements of 15.19, 13.01, and 13.53 dB in PSNR. These improvements are 1.32, 2.41, and 2.65 dB higher than using NAF.

\begin{figure*}[t]
	\begin{center}
		\begin{tabular}[t]{c} \hspace{-2.4mm}
			\includegraphics[width=0.99\textwidth]{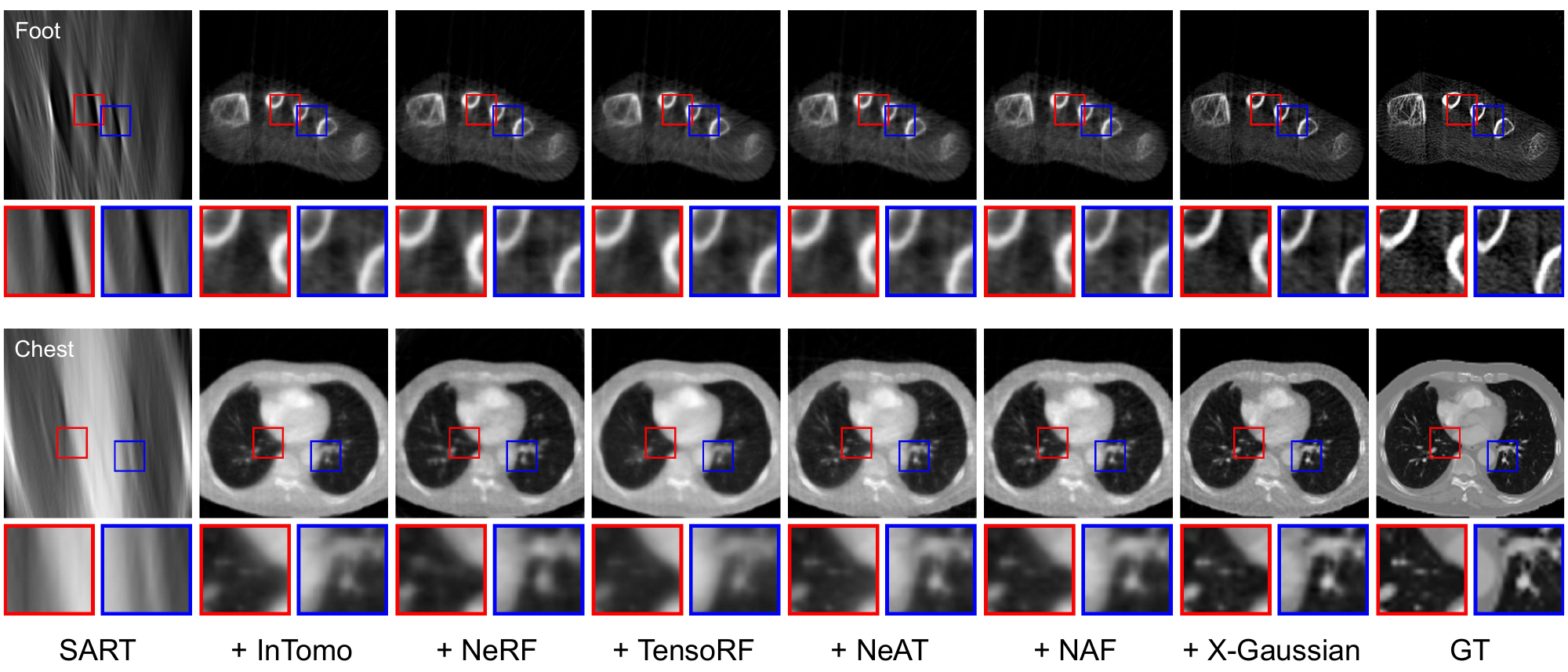}
		\end{tabular}
	\end{center}
    \vspace{-5mm}
	\caption{Visual results of sparse-view CT reconstruction on the scenes of foot and chest. NeRF-based methods and our X-Gaussian are used to create novel views for SART~\cite{sart}.}
	\label{fig:ct_compare_sart}
\end{figure*}

Figs.~\ref{fig:ct_compare_sart} shows the qualitative results of sparse-view CT reconstruction on the scenes of foot and chest. Without using rendered novel-view projections, SART fails in reconstructing the CT slices. When using SOTA X-ray NeRF-based methods to create novel views, SART produces over-smooth CT slices with blurry structural contents. On the contrary, when using our X-Gaussian to assist SART and ASD-POCS, they can reconstruct much clearer CT slices with more high-frequency textures and fine-grained structural details, such as the vessels in the chest (Fig.~\ref{fig:ct_compare_sart}). These results clearly demonstrate the potential practical values of our method on the sparse-view CT reconstruction task.

\begin{table*}[t]
	\centering
	\renewcommand{\arraystretch}{1.6}
        \caption{Results on sparse-view CT reconstruction. NeRF-based methods and our X-Gaussian are used to create novel views for FDK~\cite{fdk}, SART~\cite{sart}, and ASD-POCS~\cite{asd_pocs}.}\label{tab:ct}
	\resizebox{1\textwidth}{!}{\hspace{-4mm}
		\setlength{\tabcolsep}{0.7mm}
		\centering
		\begin{tabular}{c|cccccccccccc|cc} 
			\toprule[0.15em]
			Method & \multicolumn{2}{c}{+ None~~} & \multicolumn{2}{c}{+ InTomo~\cite{intratomo}~~~} & \multicolumn{2}{c}{+ NeRF~\cite{nerf}~~~} & \multicolumn{2}{c}{+ TensoRF~\cite{tensorf}~~~} 
			& \multicolumn{2}{c}{+ NeAT~\cite{neat}~~~}  &\multicolumn{2}{c|}{+ NAF~\cite{naf}~~~} &\multicolumn{2}{c}{+ X-Gaussian~}  \\
			Metric~&~PSNR~ & ~SSIM~ & ~PSNR~ & ~SSIM~ & ~PSNR~ & ~SSIM~ & ~PSNR~ & ~SSIM~ & ~PSNR~ & ~SSIM~ & ~PSNR~ & ~SSIM~ & ~~PSNR~~ & ~~SSIM~~ \\
			\midrule[0.15em]
			~~~FDK~  & 7.41 & 0.093 & 20.31 & 0.498 & 20.57 & 0.502 &20.61  &0.501  & 20.94 & 0.511 &21.28 &0.523 &\bf 22.60 & \bf 0.584 \\
			$\Delta$ FDK~  & 0 & 0  &12.90  &0.405  &13.16  &0.409  &13.20  &0.408  &13.52  &0.418   &13.87  &0.430   &\bf 15.19   &\bf 0.491  \\
			\midrule[0.15em]
			~~~SART~  &17.24  &0.528  & 26.28   &0.859  &26.78  &0.853  &27.06  &0.867 &27.31 &0.869 &27.84 &0.879 &\bf 30.25 &\bf 0.907 \\
			$\Delta$ SART~ &0  &0  &9.04 &0.331  &9.54  &0.325  &9.82  &0.339 &10.07 &0.341 &10.60 &0.351 &\bf 13.01 &\bf 0.379 \\
			\midrule
			~~~ASD-POCS~  &17.03  &0.525  &25.44  &0.847  &26.58  &0.857  &26.93  &0.868 &26.95  &0.865  &27.91 &0.880  &\bf 30.56  &\bf 0.926  \\
			$\Delta$ ASD-POCS~  &0  &0  &8.41  &0.322  &9.55  &0.332  &9.90  &0.343 &9.92  &0.340  &10.88 &0.355  &\bf 13.53  &\bf 0.401  \\
			\bottomrule[0.15em]
	\end{tabular}}
\end{table*}

\subsection{Ablation Study}

\subsubsection{Break-down Ablation.} 
We first conduct a break-down ablation experiment to study the effect of each proposed technique towards higher performance and faster speed. We adopt the original 3DGS~\cite{3dgs} as the baseline model and naively average the RGB channels to represent the value of radiation intensity. The results are listed in Tab.~\ref{tab:breakdown}. The baseline model yields 37.21 dB PSNR in performance. Its average training time and inference speed are 31 min 38 s and 64 fps, respectively. We can observe from Tab.~\ref{tab:breakdown} : \textbf{(i)} When using ACUI to replace the time-consuming SfM~\cite{sfm} algorithm for initialization, the training time is significantly reduced by 34\% while the performance yields an improvement of 1.66 dB. This evidence suggests that our ACUI strategy is much faster than the SfM~\cite{sfm} algorithm used in the original 3DGS and can compute more accurate camera calibration parameters for 3D Gaussians and subsequent rendering. \textbf{(ii)} Then we apply our radiative Gaussian point cloud model equipped with the proposed Differentiable Radiative Rasterization (DRR)  to replace the original RGB Gaussian point cloud model and its RGB rasterization. As analyzed in Sec.~\ref{sec:pc_model} and compared in Fig.~\ref{fig:pc_model}, the anisotropic spherical harmonics (SH) are not suitable for X-ray imaging because X-ray imaging based on penetration is isotropic. In contrast, our radiative Gaussian point cloud model can better fit the view-independent radiation intensity in 3D space. Therefore, the performance is significantly improved by 4.53 dB in PSNR. Besides, removing the computation related to the view direction from the forward and backward processes of rasterization can further accelerate the training and inference speed. Thus, the training time is reduced by 54.10\% and the inference speed is 2.1$\times$ faster.

\subsubsection{Initialization of Point Position.} 
We compare different initialization strategies for the center positions of Gaussian point clouds including random, spherical, FDK~\cite{fdk}, and cubic initialization. To be specific, Random initialization means randomly sampling points within the scanned area in 3D space. Spherical initialization uniformly samples point positions within a sphere that can completely enclose the scanned object. FDK~\cite{fdk} initialization adopts the FDK algorithm to back-project the given projections into 3D point positions. Cuboid initialization is our ACUI. Please note that we keep the computed intrinsic and extrinsic matrices the same for fair comparison between different strategies. The results are reported in Tab.~\ref{tab:init}. FDK initialization slightly outperforms our ACUI by 0.066 dB but its training time is 2.59$\times$ longer and its inference speed is 55 fps slower than ACUI. This is because the back-projection in FDK is time-consuming and initializes redundant points. The random and spherical initialization strategies yield lower PSNR and slower speed than cubic initialization.  To achieve a better trade-off, we adopt the cubic initialization, \emph{i.e.}, ACUI, which enjoys good performance, the cheapest training cost, and the fastest inference speed.

\begin{table*}[t]
	\hspace{-2mm}
        \caption{Ablation study. PSNR, SSIM, training time, and inference speed are reported.}
	\label{tab:ablations}
	\subfloat[Break-down ablation study \label{tab:breakdown}]{ 
		\scalebox{0.63}{
			\begin{tabular}{l  c c c}
				\toprule
				\rowcolor{color3} Method &~~~~3DGS~\cite{3dgs}~~ &~~~+ ACUI~~~ &~~~+ DRR~~~  \\
				\midrule
				PSNR &37.213 &38.872 &\bf 43.404 \\
				SSIM &0.9813 &0.9871 &\bf 0.9993 \\
				Train time (s) &1898 &1172 &\bf 538  \\
				Infer speed (fps) &64 &72 &\bf 148 \\
				\bottomrule
	\end{tabular}}}
	\subfloat[Ablation of point position initialization \label{tab:init}]{ 
		\scalebox{0.63}{
			\begin{tabular}{l  c c c c}
				\toprule
				\rowcolor{color3} Method &~~~Random~~~  &~~~Spherical~~~ &~~~~FDK~\cite{fdk}~~ &~~~Cubic~~~  \\
				\midrule
				PSNR &41.329 &42.837 &\bf 43.470 &43.404 \\
				SSIM &0.9942 &0.9988 &\bf 0.9993 &\bf 0.9993\\
				Train time (s) &601 &575 &1394 &\bf 538 \\
				Infer speed (fps) &112 &136 &93 &\bf 148  \\
				\bottomrule
	\end{tabular}}}
	
	\hspace{-2mm}
	\subfloat[Analysis of the number of features $N_f$ \label{tab:fea_num}]{
		\scalebox{0.64}{
			\begin{tabular}{l  c c c c c c}
				\toprule
				\rowcolor{color3} Number  &~~~~1~~~~  &~~~~2~~~~ &~~~~4~~~~ &~~~~8~~~~ &~~~~16~~~~ &~~~~32~~~~    \\
				\midrule
				PSNR &38.818 &40.205 &42.130 &42.868 &43.404 &\bf 43.417\\
				SSIM  &0.9840 &0.9931 &0.9983 &0.9991 &\bf 0.9993 &\bf 0.9993\\
				Train time (s) &\bf 511 &525 &521 &553 &538 &752\\
				Speed (fps) &153 &152 &\bf 158 &127 &148 &101\\
				\bottomrule
	\end{tabular}}}
	\subfloat[Analysis of the initialized interval $d$ \label{tab:interval}]{
		\scalebox{0.64}{
			\begin{tabular}{l  c c c c c c}
				\toprule
				\rowcolor{color3} Interval  &~~~~1~~~~  &~~~~2~~~~ &~~~~4~~~~ &~~~~8~~~~ &~~~~16~~~~ &~~~~32~~~~    \\
				\midrule
				PSNR &42.853 &42.979 &43.215 &\bf 43.404 &43.311 &43.294\\
				SSIM  &0.9989 &0.9990 &0.9992 &\bf 0.9993 &0.9992 &0.9992\\
				Train time (s) &785 &593 &545 &538 &\bf 534 &566\\
				Speed (fps) &86 &94 &114 &\bf 148 &135 &97\\
				\bottomrule
	\end{tabular}}}
    \vspace{-4mm}
\end{table*}

\subsubsection{Parameter Analysis.} 
We conduct parameter analysis of the number of features $N_f$ and the sampling interval $d$. The results are shown in Tab.~\ref{tab:fea_num} and Tab.~\ref{tab:interval}. 

In Tab.~\ref{tab:fea_num}, \textbf{(i)} When increasing $N_f$, the performance gradually improves but the magnitude of the improvement decreases. In particular, $N_f = 32$ achieves the best results of 43.42 dB in PSNR. $N_f = 16$ achieves on-par results with $N_f = 32$, only 0.013 dB lower. \textbf{(ii)} We notice that the training time and inference speed do not change monotonically. This is because the Gaussian point clouds with various feature dimensions have different representing ability and computational complexity. The number of final 3D Gaussians after training also varies substantially. When $N_f = 16$, the training time reaches a local minimum and the inference speed is at its local maximum. Hence, we eventually adopt $N_f = 16$ to reach a more optimal balance between performance and speed. 

In Tab.~\ref{tab:interval}, the best reconstruction performance and the fastest inference speed are achieved at $d = 8$. The training time (538 s) at $d = 8$ is almost the same as the shortest one (534 s) at $d = 16$. Thus, we eventually set $d$ to 8.

\subsubsection{Convergence Analysis.} 
\begin{figure}[t] 
\begin{minipage}[t]{0.545\textwidth}%
    \noindent \begin{center}
        \includegraphics[width=1.0\columnwidth]{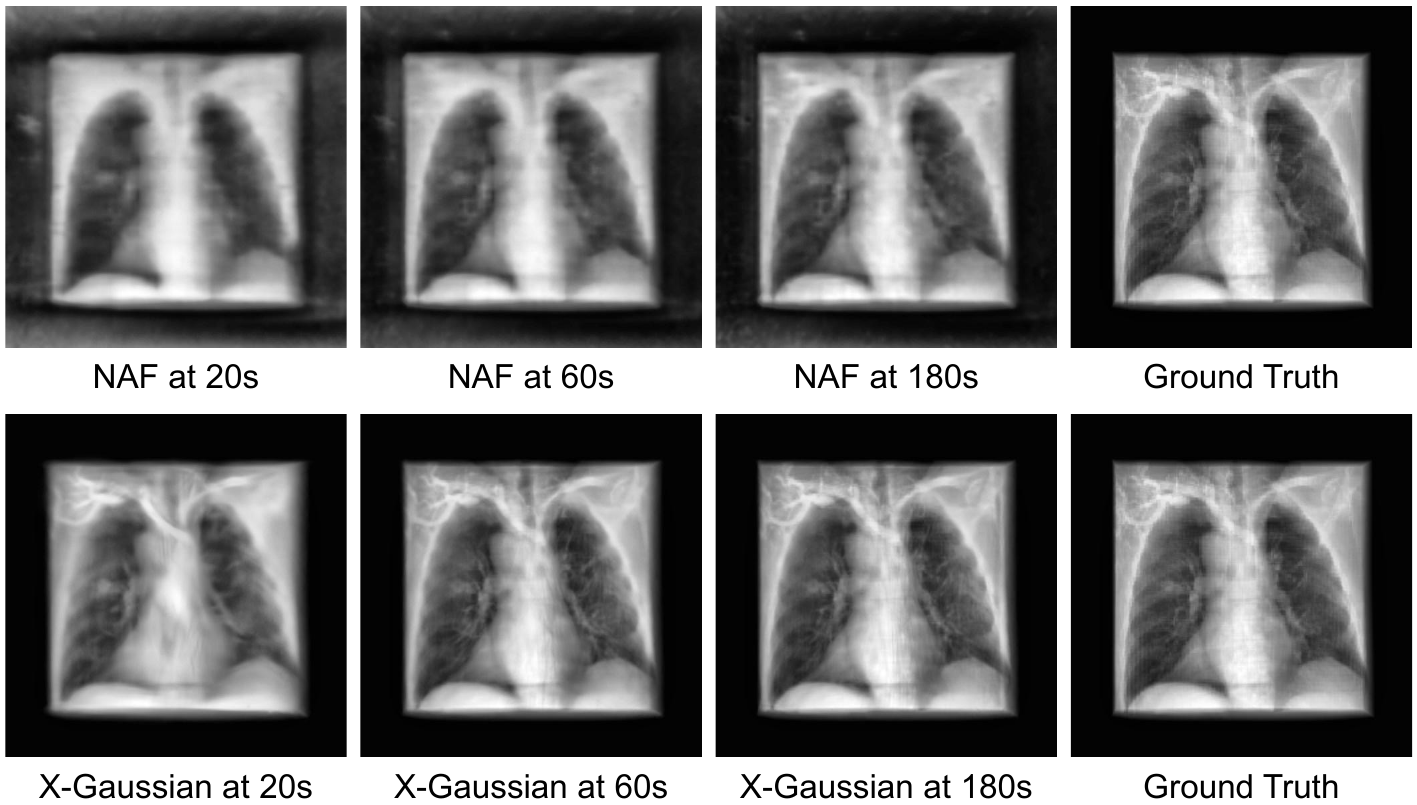} 
        \caption{Convergence analysis of NAF~\cite{naf} \emph{vs.} our X-Gaussian. We visualize the rendered projections at 20s, 60s, and 180s of training. Our X-Gaussian shows faster and better convergence.}
        \label{fig:visual_analysis_proj} 
        \par\end{center}%
\end{minipage}
    \hspace{0.3mm}
\begin{minipage}[t]{0.412\textwidth}%
    \noindent \begin{center}
        \includegraphics[width=1.0\columnwidth]{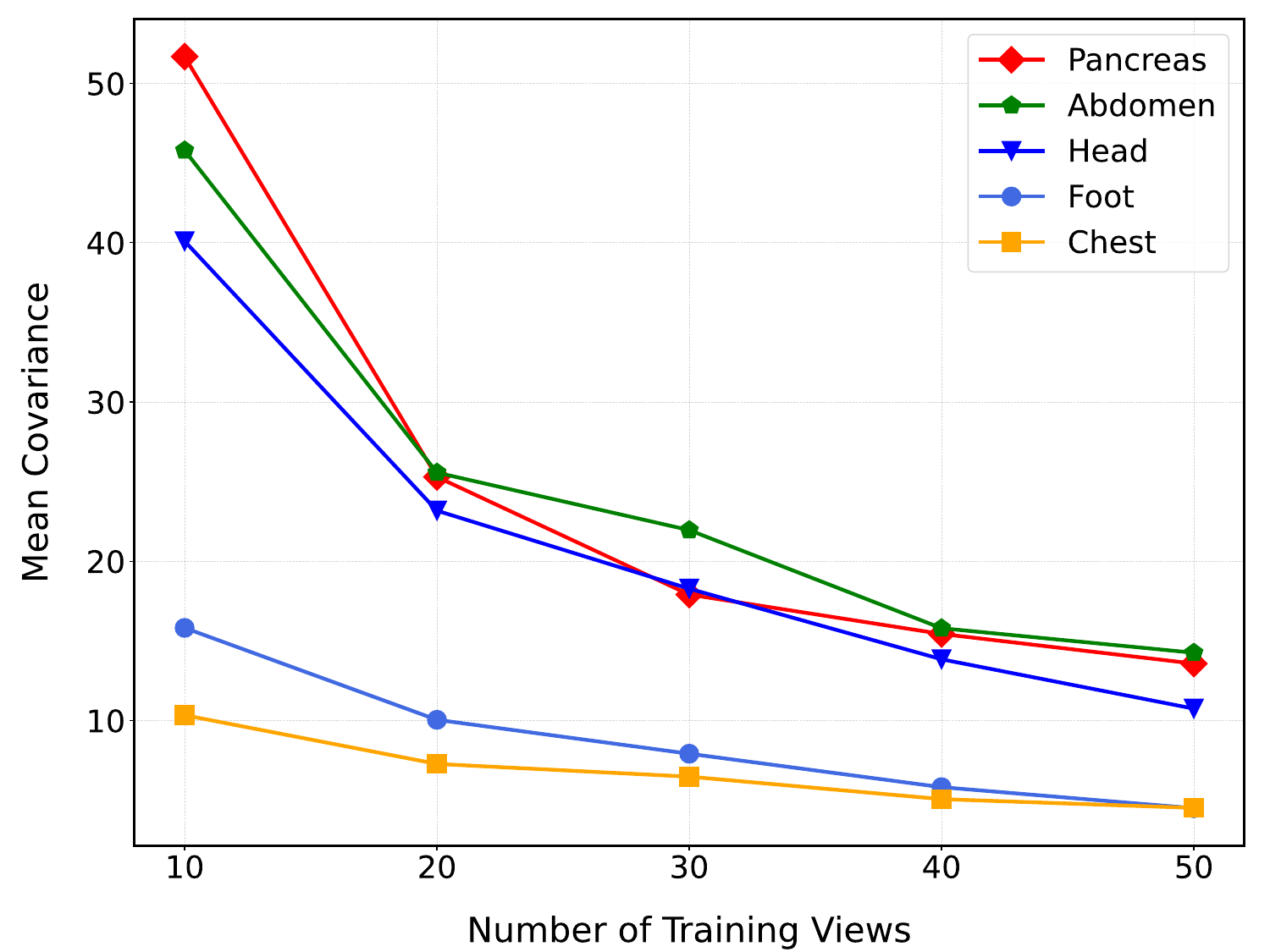} 
        \caption{Analysis of covariance. The mean covariance of 3D Gaussians on different scenes decreases when the number of training views increases.}
        \label{fig:covariance} 
        \par\end{center}%
\end{minipage}
\end{figure}

We conduct two visual analyses to compare the convergence between our X-Gaussian and original 3DGS~\cite{3dgs} in Fig.~\ref{fig:analysis_3dgs}, and between X-Gaussian and the SOTA X-ray NeRF-based method NAF~\cite{naf} in Fig.~\ref{fig:visual_analysis_proj}.

Specifically, we adopt the same ACUI strategy for the original 3DGS to focus on comparing the Gaussian point cloud model and the rasterization. For fairness, we train 3DGS and our X-Gaussian on the scene of foot with the same settings and visualize the positions and opacities of Gaussian point clouds at the 100-$th$, 1000-$th$, 5000-$th$, 10000-$th$, and 20000-$th$ iterations of the training process in Fig.~\ref{fig:analysis_3dgs}. We also visualize the CT volume of foot as a reference. As can be seen that 3DGS with RGB rasterization converges slowly and suffers from more noisy point clouds. Plus, the final trained model of 3DGS at the 20000-$th$ iteration contains more redundant Gaussians that are unnecessary to represent the 3D structure of the foot, which reduces the model's inference speed. In contrast, our X-Gaussian equipped with the proposed DRR shows faster and better convergence. In particular, as early as the 1000-$th$ iteration, our radiative Gaussian point clouds have essentially formed the basic shape of the foot. Besides, the final trained X-Gaussian at the 20000-$th$ iteration can better represent the 3D geometry and more accurate structural contents than the original 3DGS.

\vspace{-3mm}

Additionally, in Fig.~\ref{fig:visual_analysis_proj}, we visualize the rendered novel projections with the same angle $\phi$ of NAF and our X-Gaussian at 20s, 60s, and 180s of the training process on the scene of chest. NAF produces blurry images with severe noises on the background regions within the first three minutes of the training. In contrast, our X-Gaussian can reconstruct clearer structural details like the ribs and blood vessels with cleaner background of the chest at the first minute of the training.

\subsubsection{Analysis of Covariance.} We study how the shape of the 3D Gaussian point cloud changes with the number of training views in Fig.~\ref{fig:covariance}. As the number of training views increases, the mean covariance of 3D Gaussians that control the size of Gaussian point clouds decreases. This indicates that the 3D Gaussian point clouds gradually change from coarse to fine, thereby being more capable of representing fine-grained structures, such as a small tumor in the abdomen.

\section{Conclusion}
In this paper, we propose the first 3DGS-based framework, X-Gaussian, for X-ray 3D reconstruction. Firstly, we design a radiative Gaussian point cloud model based on X-ray imaging properties. This model excludes the influence of view direction when fitting radiation intensity. For this model, we develop a GPU-friendly differentiable radiative rasterization CUDA kernel that renders projections at a faster speed than RGB rasterization. Secondly, we formulate a principled RGVQ algorithm to construct a continuous attenuation field from the original radiative point attributes and reconstruct CT volumes by through CUDA-based parallel query-
ing, extending conventional 3DGS beyond image rendering. Thirdly, we customize an SfM-free strategy, ACUI, for fast initialization. Experiments show that our X-Gaussian outperforms SOTA methods by 6.5 dB while enjoying 73$\times$ faster inference speed and only requiring 15\% of training time on X-ray NVS.  In addition, X-Gaussian
surpasses SOTA methods by 0.40 dB and 4.34 dB in 
PSNR on synthetic and real CT reconstruction.

\subsubsection{Acknowledgements:} This work was supported by the Lustgarten Foundation for Pancreatic Cancer Research and the Patrick J. McGovern Foundation Award.

\bibliographystyle{splncs04}
\bibliography{reference}
\end{document}